# The performance of microwave photonic signal processors based on microcombs with different input signal waveforms

**David J. Moss**

Optical Sciences Center, Swinburne University of Technology, Hawthorn, VIC 3122, Australia;
\* Correspondence: dmoss@swin.edu.au

**Abstract:** Microwave photonic (MWP) signal processors, which process microwave signals based on photonic technologies, bring advantages intrinsic to photonics such as low loss, large processing bandwidth, and strong immunity to electromagnetic interference. Optical microcombs can offer a large number of wavelength channels and compact device footprints, which make them powerful multi-wavelength sources for MWP signal processors to realize a variety of processing functions. In this paper, we experimentally demonstrate the capability of microcomb-based MWP signal processors to handle diverse input signal waveforms. In addition, we quantify the processing accuracy for different input signal waveforms, including Gaussian, triangle, parabolic, super Gaussian, and nearly square waveforms. Finally, we analyze the factors contributing to the difference in the processing accuracy among the different input waveforms, and our theoretical analysis well elucidates the experimental results. These results provide a guidance for microcomb-based MWP signal processors when processing microwave signals of various waveforms.

## 1. Introduction

Microwave signal processors have found wide applications in telecommunication and radar systems [1-4]. Traditional microwave signal processors relying on electronic devices exhibit significant loss and strong crosstalk when handling high-frequency microwave signals, which make them suffer from limited operation bandwidths. To overcome this restriction, microwave photonic (MWP) signal processors that perform signal processing functions based on MWP technologies have attracted great interests [3-6].

A variety of MWP signal processors have been demonstrated by exploiting different optical filtering modules to process microwave signals modulated onto a single optical carrier [7-17]. Although these approaches feature high performance in achieving specific processing functions, they face limitations in their reconfigurability to realize diverse processing functions based on a single system. On the contrary, in MWP signal processors implemented based on the transversal filter structure [18], input microwave signals are modulated onto multiple optical carriers with adjustable time delays and tap weights before summing via photodetection. This enables a high reconfigurability to achieve various processing functions without changing any hardware [2, 18].

For MWP signal processors implemented by the transversal filter systems, a large number of taps, or the wavelength channels provided by multi-wavelength optical sources, is required to improve their performance. Compared to other multi-wavelength optical sources, such as discrete laser arrays [19-21], fibre Bragg grating arrays [22-24], laser frequency combs generated by electro-optic (EO) modulation [25-27], and mode-locked fiber lasers [28, 29], optical microcombs can provide a large number of wavelength channels by using compact micro-scale resonators [3, 4, 30]. They are also with the ability to offer broad Nyquist zones, which allow for large processing bandwidths [4, 31, 32]. With these advantages, a variety of signal processing functions have been successfully demonstrated using microcomb-based MWP signal processors, such as differentiation [33], integration [34], Hilbert transform [35], arbitrary waveform generation [36], and convolutional processing [37, 38].

Although a range of signal processing functions have been realized, they only used Gaussian input waveforms for demonstrations, while the ability to handle various input signal waveforms is essential for practical applications. In this paper, we experimentally demonstrate the capability of microcomb-based MWP signal processors for dealing with various input signal waveforms. We investigate the processing accuracy of different input waveforms, including Gaussian, triangle, parabolic, super Gaussian, and nearly square waveforms. We also perform theoretical analysis and discuss the reasons for the difference in the processing accuracy among the different input waveforms. These results offer a valuable guide for microcomb-based MWP signal processors to handle microwave signals with different waveforms.

## 2. Microcomb-based MWP signal processors

MWP signal processors based on the transversal filter are implemented based on MWP technologies, which can overcome the electrical bandwidth bottleneck by providing a substantially increased processing bandwidth. A MWP transversal signal processor has a high reconfigurability in terms of its spectral transfer function, which can be expressed as [4]

$$H(\omega) = \sum_{n=0}^{M-1} a_n e^{-j\omega n \Delta T}, \quad (1)$$

where $\omega$ is the angular frequency of the input microwave signal to be processed, $M$ is the tap number, $a_n$ ($n$ = 0, 1, 2, …, $M$-1) is the tap coefficient of the $n^{th}$ tap, and $\Delta T$ is the time delay between adjacent taps. By properly designing the various tap coefficients $a_n$ ($n$ = 0, 1, 2, …, $M$-1), different signal processing functions can be realized by using a single system without changing the hardware.

**Figure 1** shows the schematic of a microcomb-based MWP signal processor. An optical microcomb is used to generate multiple wavelength channels that act as discrete taps for the transversal signal processor. The generated optical microcomb is spectrally shaped according to the designed tap coefficients $a_n$ ($n$ = 0, 1, 2, …, $M$-1). Next, all of the wavelength channels of the shaped optical microcomb are imprinted with the input microwave signal via an electro-optic modulator (IM), leading to the generation of multiple microwave replicas. Following this, the modulated optical signals transmit through a dispersive medium to introduce time delays $\Delta T$, which progressively separate the microwave replicas. Finally, the delayed replicas are summed upon photodetection via a photodetector.

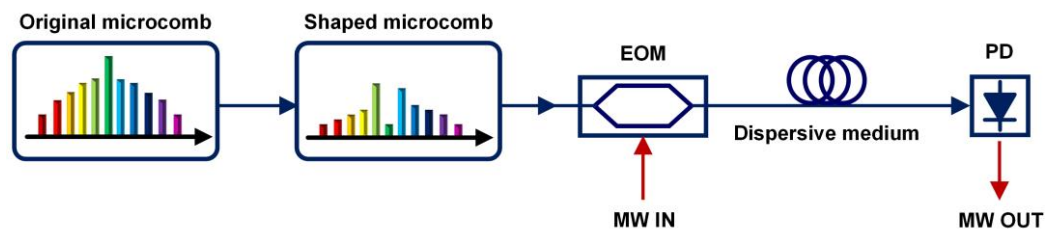

**Figure 1.** Schematic diagram of a microcomb-based microwave photonic (MWP) signal processor. EOM: electro-optic modulator. MW: microwave. PD: photodetector.

## 3. Experimental results

In our experimental demonstration, we implemented the microcomb-based MWP signal processors based on the setup shown in **Figure 2**, which consisted of a microcomb generation module and a transversal signal processing module. In the microcomb generation module, the optical microcomb was generated by a microring resonator (MRR) made from high-index doped silica glass [39-106]. The high-index doped silica glass offers attractive material properties for microcomb generation, including ultra-low linear loss (~0.06 dB/cm), a moderate nonlinear parameter (~233 $W^{-1} \cdot km^{-1}$), and a negligible nonlinear loss even at extremely high intensities (~25 $GW \cdot cm^{-2}$) [39-106]. The MRR had a quality factor of ~1.5 × $10^6$. A continuous-wave (CW) light was amplified to ~32.1 dBm by an erbium-doped fibre amplifier (EDFA) and used to pump the MRR. The polarization of the CW pump was adjusted to TE polarization, which aligned with a TE-polarized resonance of the MRR at ~1551.23 nm. When the pump power of the CW laser was sufficient high and its wavelength was swept across the MRR's resonance at ~1551.23 nm, optical parametric oscillation occurred, resulting in the generation of a palm-like soliton crystal microcomb [40, 78-106], as shown in **Figure 3(a)**. The MRR was designed to have a radius of ~592 μm, which corresponded to a comb spacing of ~0.4 nm or ~49 GHz. In our experimental demonstration, 20 comb lines were employed to as discrete taps. The initially generated microcomb exhibited non-uniform power distributions among the comb lines and so it was shaped by the first waveshaper (WS1, Finisar) to flatten the comb lines. This was done to achieve a higher signal-to-noise ratio and reduce the required loss control range for the second waveshaper in the transversal signal processing module, which further shaped the comb lines according to the designed tap coefficients.

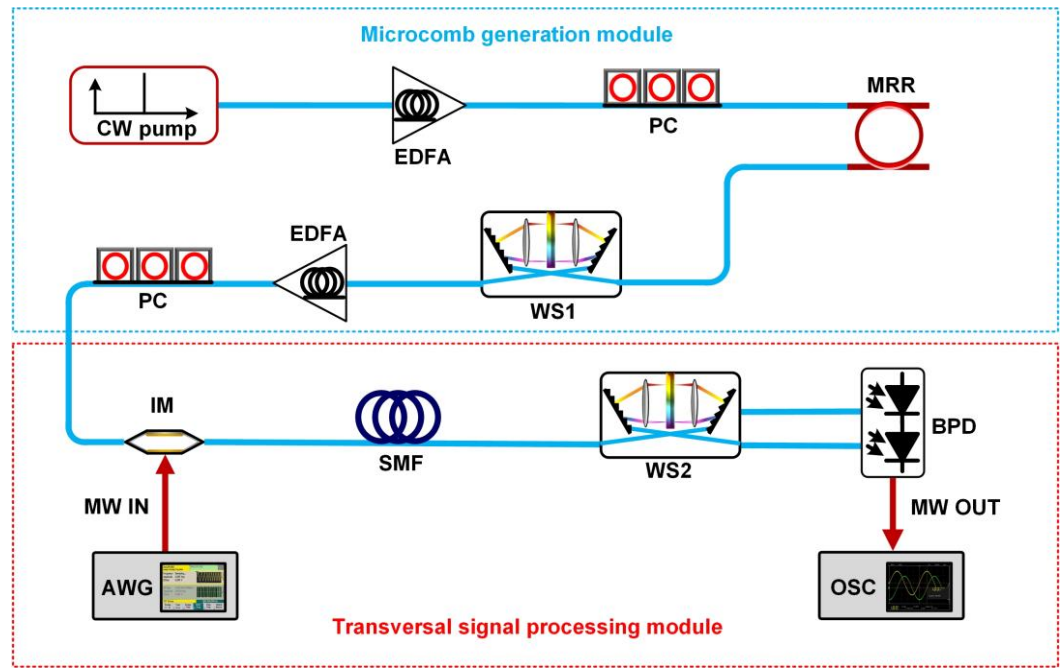

**Figure 2.** Experimental schematic of a microcomb-based MWP signal processor. CW pump: continuous-wave pump. EDFA: erbium-doped fibre amplifier. PC: polarization controller. MRR: microring resonator. WS: wave shaper. IM: intensity modulator. MW: microwave. BPD: balanced photodetector. SMF: single-mode fibre. AWG: arbitrary waveform generator. OSC: oscilloscope.

In the transversal signal processing module, the shaped microcomb was modulated by the input microwave signal via an intensity modulator (IM) (iXblue). The input microwave signal was multicast onto different wavelength channels, resulting in the generation of multiple microwave replicas. Next, the microwave replicas were transmitted through a spool of single mode fibre (SMF), which served as the dispersive medium that introduced a time delay between adjacent wavelength channels, *i.e.*, $\Delta T$ in **Eq. (1)**. The time delay $\Delta T$ can be further expressed as [4]

$$\Delta T = L \times D_2 \times \Delta \lambda \qquad (2)$$

where $L$ is the fibre length, $D_2$ is the second-order dispersion parameter, and $\Delta\lambda$ is the comb spacing. In our experiments, these parameters were $L$ = ~5.124 km, $D_2$ = ~17.4 ps/nm/km, and $\Delta\lambda$ = ~0.4 nm, which resulted in a time delay $\Delta T$ = ~ 35.7 ps.

After passing the dispersive medium, the comb lines were spectrally shaped by the second waveshaper (WS2, Finisar) according to the designed tap coefficients $a_n$ ($n$ = 0, 1, 2, …, $M$-1). Finally, the delayed microwave replicas were summed upon photodetection via a balanced photodetector (BPD, Finisar). The BPD separated the wavelength channels into two categories according to the sign of tap coefficients, achieving both positive and negative tap coefficients.

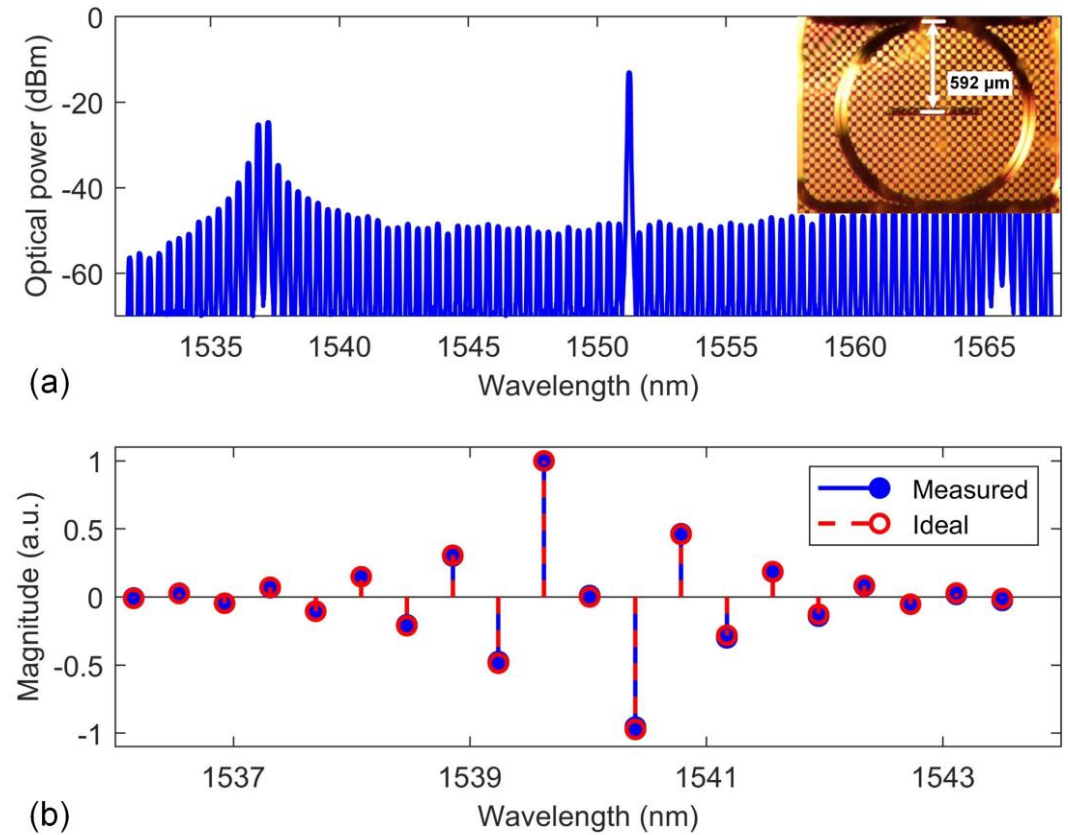

**Figure 3.** (a) Optical spectrum of soliton crystal microcomb generated by a MRR made from high-index doped silica glass. Inset shows a microscope image of the MRR. (b) Ideal and measured tap coefficients after optical spectral shaping.

We took the first-order differentiation as an example to investigate the influence of different input signal waveforms on the processing accuracy of microcomb-based MWP signal processors. The spectral transfer function of the first-order differentiation can be described by [3]

$$H(\omega) = j\omega, \tag{3}$$

where $j = \sqrt{-1}$, and $\omega$ is the angular frequency. The ideal tap coefficients were calculated by performing an inverse Fourier transform of **Eq. (3),** and the results is shown in **Figure 3(b)**. For comparison, the measured tap coefficients after spectral shaping of the comb lines are also shown. As can be seen, the measured tap coefficients closely matched with the ideal tap coefficients, indicating the achievement of effective spectral shaping.

We selected five different temporal waveforms for the input microwave signal, including Gaussian, triangle, parabolic, super Gaussian, and nearly square waveforms. The input microwave signals were generated by an arbitrary waveform generator (AWG, Keysight). According to the Nyquist sampling theorem, the sampling rate of a continuous-time bandwidth-limited signal needs to exceed twice its maximum frequency component to avoid aliasing. This constraint sets an upper threshold for the bandwidth of the input microwave signal to be processed, which should not surpass half of the microcomb's comb spacing, *i.e.*, ~24.5 GHz. On the other hand, the FSR of the RF spectral response (FSR$_{RF}$) of the differentiator was inversely related to the time delay (**Eq. (2)**) $1/\Delta T$ = ~28 GHz. Therefore, the operation bandwidth of the signal processor is given by $f_{OB}$ = ½ FSR$_{RF}$ = ~14 GHz, which sets another limitation for the maximum bandwidth of the input microwave signal. Considering these factors, in our experiments we employed input microwave signals with a full width at half maximum (FWHM) of ~0.2 ns (**Figure 4(a)**) and the primary frequency components resided within 14 GHz.

The signal processing results are shown in **Figure 4(b)**, which were measured by a high-speed real-time oscilloscope (OSC, Keysight). The theoretical outputs are also shown for comparison, which were calculated based on **Eqs. (1) – (3)**. To facilitate a fair comparison, we used the recorded waveforms generated by the AWG as the input signal waveforms to calculate the theoretical outputs. As can be seen, all the measured outputs match with their corresponding theoretical outputs. Nevertheless, different input waveforms exhibit differences in the discrepancies between them. The Gaussian input waveform shows the lowest discrepancies, whereas the nearly square waveform displays the highest.

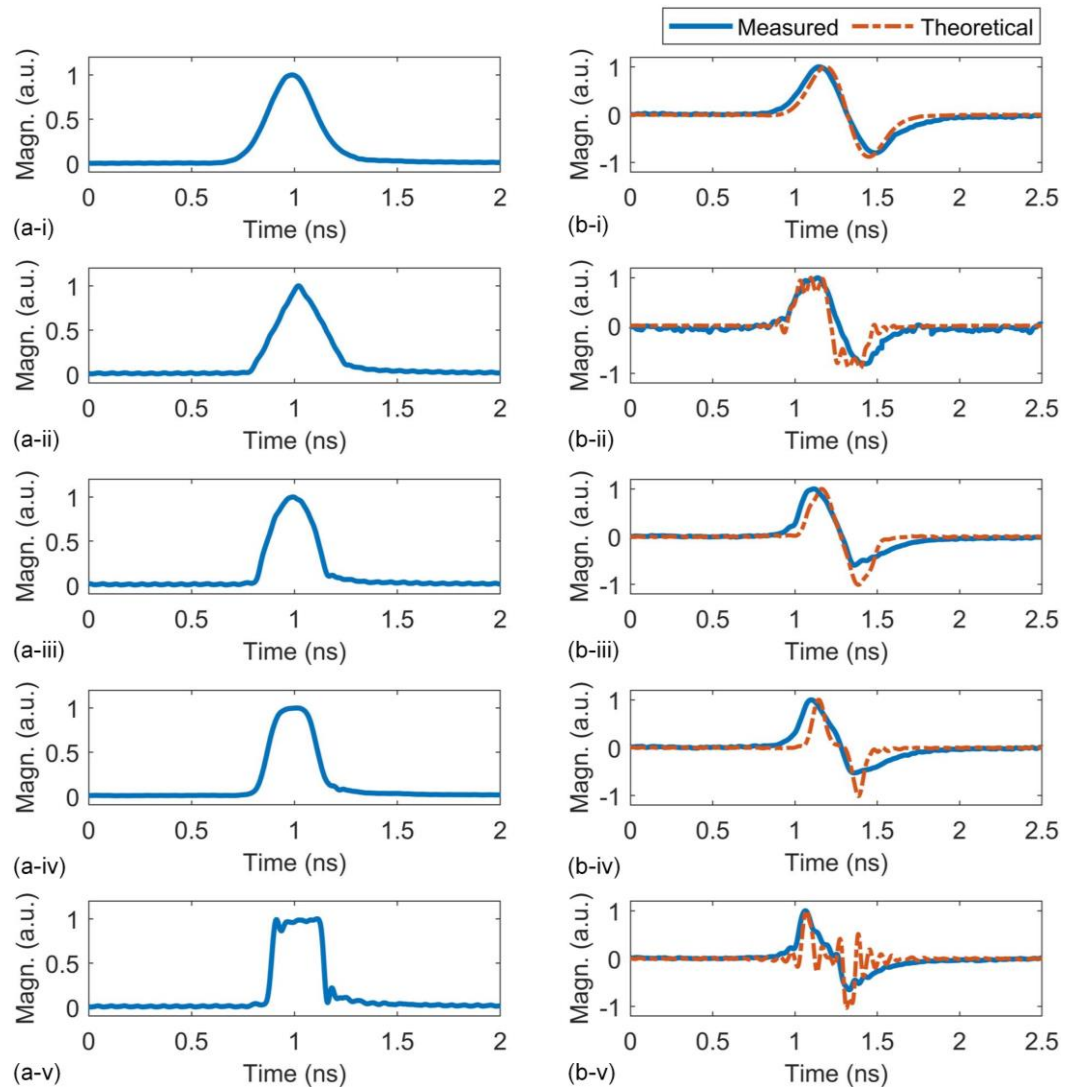

**Figure 4.** (a) Measured input microwave signal waveforms of (i) Gaussian, (ii) triangle, (iii) parabolic, (iv) super Gaussian, and (v) nearly square waveforms with full width at half maximum (FWHM) of ~0.2 ns. (b) Measured output waveforms from the MWP signal processor that performs first-order differentiation. The theoretical output results are also shown for comparison.

To quantify the processing accuracy of the processing results, the concept of root mean square error (RMSE) is introduced, which is defined as [30]

$$\text{RMSE} = \sqrt{\sum_{i=1}^{k} \frac{(Y_i - y_i)^2}{k}} \quad (4)$$

where $Y_1, Y_2, \ldots, Y_k$ are the values of theoretical processing results, $y_1, y_2, \ldots, y_k$ are values of measured output waveforms.

**Figure 5(a)** shows the RMSEs between the measured output waveforms and the theoretical processing results for different input signal waveforms. The Gaussian and nearly square waveforms have the lowest and highest RMSE values, showing agreement with the results in **Figure 4(b)**.

To analyze the reason for the differences in the processing accuracy for different waveforms, we further plot the amplitude frequency response of the processor and a theoretical differentiator in **Figure 5(b)**, together with the spectra of input signals with different waveforms. It can be seen that the deviations between the response of the transversal signal processor and the theoretical differentiator become more significant in the high-frequency range. On the other hand, the nearly square waveform contains greater high-frequency components than other waveforms, which results in a reduction in its processing accuracy. In contrast, the Gaussian waveform has the least high-frequency components, enabling the highest level of processing accuracy.

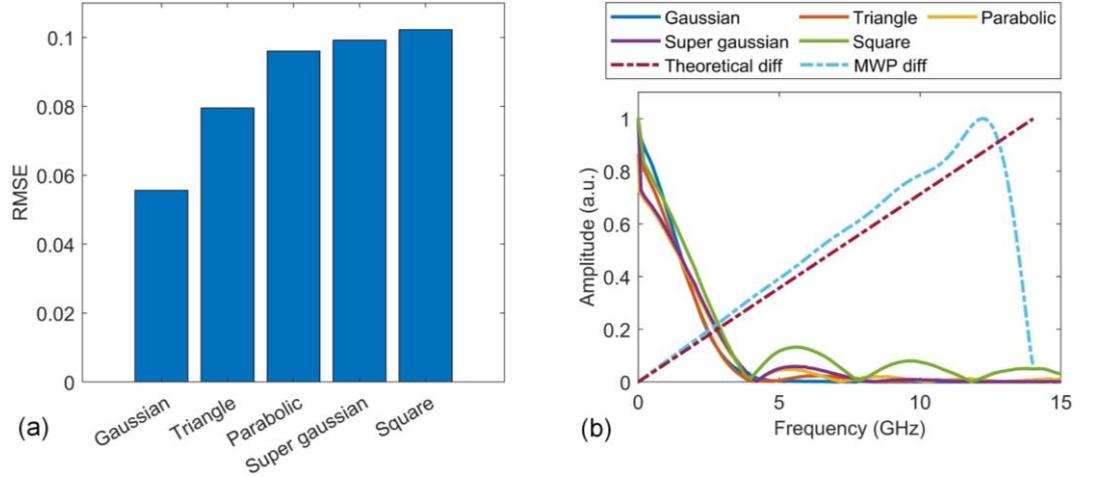

**Figure 5.** (a) RMSEs between theoretical differentiation results and the processor's output waveforms for different input microwave signal waveforms in **Figure 4**. (b) Amplitude frequency response of theoretical differentiation and the processor, together with the amplitude spectra of different input microwave signals including with Gaussian, triangle, parabolic, super Gaussian, and nearly square waveforms shown in **Figure 4(a)**.

Based on the above results, it can be seen that the processing accuracy varies for different input signal waveforms, even when performing the same processing function. The processing accuracy improves when there is better overlap between the high-intensity frequency components of the input signal and the low-error region of the MWP processor's response spectrum. These results have implications for a wide range of linear and nonlinear photonic devices. [107-152]

## 4. Conclusion

In summary, we experimentally demonstrate that microcomb-based MWP signal processors are capable of processing microwave signals with different temporal waveforms. We characterize the processing accuracy for different input signal waveforms, including Gaussian, triangle, parabolic, super Gaussian, and nearly square waveforms. We find that the difference in the processing accuracy for various input waveforms is mainly resulting from the difference in their frequency components, as well as their overlap with the processor's frequency response that exhibit different degrees of deviation from the ideal response. These results provide a useful guidance for microcomb-based MWP signal processors to process microwave signals with various waveforms.

**Institutional Review Board Statement:** Not applicable.

**Informed Consent Statement:** Not applicable.

**Data Availability Statement:** Not applicable.

**Conflicts of Interest:** The authors declare no conflict of interest.